\documentclass[aps,twocolumn, amsfonts,amsmath,tightenlines,showpacs,nofootinbib]{revtex4}
\usepackage{epsf}

\def\be{\begin{equation}}
\def\ee{\end{equation}}
\def\beq{\begin{equation}}
\def\eeq{\end{equation}}
\def\bea{\begin{eqnarray}}
\def\eea{\end{eqnarray}}
\def\bml{\begin{subequations}}
\def\blea{\bml\begin{eqnarray}}
\def\elea{\end{eqnarray}\end{subequations}}

\begin{document}

\title{Non-linear dynamics of cosmic strings with non-scaling loops}

\author{Vitaly Vanchurin}

\email{vanchurin@stanford.edu}

\affiliation{Department of Physics, Stanford University, Stanford, CA 94305}

\begin{abstract}

At early stages the dynamics of cosmic string networks is expected to be influenced by an excessive production of small loops at the scales of initial conditions $l_{min}$. To understand the late time behavior we propose a very simple analytical model of strings with a non-scaling population of loops. The complicated non-linear dynamics is described by only a single parameter $N \sim 2/(1-C(l_{min}))$ where $C(l)$ is a correlation function of the string tangent vectors. The model predicts an appearance of two new length scales: the coherence length $\xi \sim t/N^2$ and the cross-correlation length $\chi \sim t/N$. At the onset of evolution $N\sim10$ and at late times $N$ is expected to grow logarithmically due to cosmological stretching and emission of small loops. The very late time evolution might be modified further when the gravitational back-reaction scale grows larger than $l_{min}$.

\end{abstract}


\maketitle

\section{Introduction}

Cosmic strings are predicted by many models of symmetry breaking phase transitions \cite{Kibble} and give rise to very distinct and detectable signatures such as gravitational lensing \cite{Vilenkin, book}, CMB non-Gaussianities \cite{FRSB,TNSYYS,Vanchurin2, Hindmarsh, RS}, gravitational waves \cite{DV} and ultra-high energy cosmic rays \cite{BV, BKV}. More recently, it was also realized that cosmic super-strings are usually formed at the end of brane inflation \cite{Tye,Dvali,Polchinski} which opens a possibility of testing the models of string theory in the cosmological settings. 

In the early work on cosmic strings, it was expected \cite{AV81,Kibble85} that the typical length of closed loops scales linearly with time, but some old numerical simulations showed no evidence of such scaling. After more than twenty years of numerical studies \cite{BB,AS, BB91, Hindmarsh, RSB, VOV, VOV2, OV, MS} the issue of string loops is still a subject of an ongoing debate. Some groups argue that the loops are much smaller than horizon \cite{BB91, RSB, VHS, HSB}, when others claim that the loops are typically produced at the near horizon scales \cite{MS, VOV, VOV2, OV}. Clearly the final verdict has not yet been reached. 

In addition to numerical studies a number of very interesting analytical models were proposed \cite{CK, ACK, MartinsShellard, DPR, Rocha, Vanchurin, CK2}, some of which claim remarkable agreements with numerical simulations  \cite{DPR}. However, the downside of many models is that they often contain phenomenological parameters, such as the mean velocity of strings, which are not derived from first principles. It is always left to numerical simulations to fix the unknown parameters and functions, which might be a dangerous path given that the dynamical range of the simulation is still very short. Moreover all of these models do not include the non-linear effects of loop production in a self-cosictent way. 

In this paper we will make a first step to develop a self-consistent dynamical model by including a back-reaction from non-scaling loops. The main assumptions are:\\
1) non-scaling loops are predominantly produced at the scales of the initial correlation length $\sim l_{min}$. \\
2) large scale $\sim t$ inter-commutations do not significantly affect the statistical properties on smaller scales $\sim l_{min}$.\\
3) in the comoving coordinates $l_{min}$ remains constant, when all other length scales grow linearly with time.\\
These assumption are motivated by many numerical simulations \cite{RSB, VOV, VOV2, OV, MS} as well as by analytical results \cite{DPR, PR, Vanchurin, Vanchurin3}.

The paper is organized as follows. In the second section we derive the most important properties of the model under assumption of slowly changing power on the smallest scales, and in the third section we show that this power decays only logarithmically. The main results are discussed in the conclusion.

\section{Non-scaling model}

One common feature of many numerical simulation \cite{MS, VOV, VOV2, OV} is an excessive production of very small loops at the sizes of the initial correlation length $l_{min}$. This production does not go away quickly, and even at late times when strings become relatively smooth the small loops are abundantly produced
\be
\langle l(t) \rangle \sim l_{min}.
\ee
At first this seems very counterintuitive given that long strings exhibit full scaling
\be
d(t) \propto \zeta (t) \propto t, 
\ee 
where $d$ is the inter-string distance and $\zeta$ is the correlation length. However, one can show that the smallest wiggles of size $l_{min}$ are very likely to form loops on a passage through each other if they form a cusp \cite{DPR, Vanchurin3}. 

Nambu-Goto evolution of strings is usually described by decomposition of a position three vector into right and left moving waves
\be
\bold{x}(\sigma, t) = \frac{\bold{a}(\sigma- t) +\bold{b}(\sigma + t) }{2}
\ee
with condition $|\bold{a}'|=|\bold{b}'|=1$, where prime denotes a derivative with respect to $\sigma$. If $\bold{a}'$ and $-\bold{b}'$ curves, corresponding to the two wiggles, intersect on a unit sphere, then they would annihilate with a significant probability \cite{DPR, Vanchurin3}. Therefore, if we choose two opposite-moving wiggles of size $l_{min}$ at random then the probability $p$ for them to intersect on a unit sphere can be calculated from the mean amplitude of wiggles
\be
P(l) \equiv \sqrt{2-2 C(l)}
\ee
 where 
 \be
 C(l) \equiv \langle \bold{a}'(0) \bold{a}'(l)\rangle
 \label{eq:correlation}
 \ee
 is a two-point correlation function and thus,
 \be
 p \sim \frac{\pi P(l_{min})^2}{4 \pi} = \frac{1-C(l_{min})}{2}.
 \ee
 
 For the following analysis it will be useful to assume that all $l_{min}$ size wiggles can be grouped into 
\be
N \equiv \frac{1}{p} \sim \frac{2}{1-C(l_{min})}
\ee
 equivalence classes (we call simply {\it directions}\footnote{Strictly speaking each class is not a single directions, but a set of directions described by area $\sim \pi P(l_{min})^2$ on a unit sphere.}) such that any pair (one left moving wiggle and one right moving wiggle) from the same equivalence class would form a loop on the passage through each other and any pair from distinct classes would pass through without interactions. Of course the division into equivalence classes is not precise, but it provides us with an intuitive picture of a process under consideration.

To understand the evolution of the entire network of strings it is convenient to introduce a new scale, we call the cross-correlation length\footnote{The existence of cross-correlations is not new (See Ref. \cite{CK}).}
\be
\chi(t) = k t,
\ee
which describes a distance along the string on which correlations between opposite moving segment had been established. More precisely, $\chi$ is an average distance along string on which the two sets: directions of all left-moving wave and directions of all right-moving segments, are disjoint. The cross-correlation length $\chi(t)$ should be smaller than time $t$ due to causality, but larger than the coherence length
\be
\xi(t) = c t,
\ee
which is defined as an average distance along the string where the direction does not change. (Note that both scales: coherence length and cross-correlation length are taken to be linear functions of time due to our original assumptions listed in the introduction.) If the left and right moving waves were completely random then one would have to make $\sim N$ comparisons before an arbitrary pair of opposite moving directions could coincide. This would make $\chi(t) \sim \xi(t) \sqrt{N}$, but because of the cross-correlations, as we will argue below, $\chi(t) \sim \xi(t) N$.

If the decay process of a long piece of string $L(t)$ into small loops is effective only on distances comparable to cross-correlation length then
\be
\frac{d L(t)}{dt} = -\frac{L(t)}{k t N},
\label{eq:small_loops_decay}
\ee
which reflects that a given wiggle has to move a distance $ \sim \chi(t)$ before meeting a "truly" random opposite moving wiggle with whom he can annihilate with probability $1/N$. This also tells us that an arbitrary segment has to move a distance $\sim N \chi(t)$ in order to form a non-scaling loop with a significant probability $\sim O(1)$. The solution of the differential equation (\ref{eq:small_loops_decay}) is given by 
\be
L\propto t^{-\frac{1}{k N}}.
\label{eq:decay_solution}
\ee

Consider a very large loop of size  
\be
L(t_1) \sim k t_2 \left ( \frac{t_1}{t_2} \right )^{-\frac{1}{k N}}
\ee
at time $t=t_1$ and evolve it forward in time until time $t_2$ with only non-scaling loops production turned on. Until the time when its length is larger than the cross-correlation length, the evolution is described on average by (\ref{eq:small_loops_decay}) with solution (\ref{eq:decay_solution}). This implies that at time $t_2$ the remaining length is equal to the cross-correlation length 
\be
L(t_2) \sim k  t_2,
\ee
and afterwards  $t \gtrsim t_2$ the decay process stops $L(t) \sim L(t_2)$.

The total number of coherence segments $\xi(t_1)=c t_1$ on the original piece of string is $\sim \frac{L(t_1)}{c t_1}$, or on average $\frac{L(t_1)}{N c t_1}$ in each of $N$ directions. Due to statistical fluctuations this number for left - and right-moving waves is not the same and under assumption of Poission statistics $\sim N \sqrt{\frac{L(t_1)}{Nct_1}}$ of the segments would survive after the cross-correlations are established on the entire piece of string $L(t_2)$. In other words
\be
N \sqrt{\frac{k t_2 \left ( \frac{t_1}{t_2} \right )^{-\frac{1}{k N}}}{Nc t_1}}  c t_1 \approx k t_2
\ee
or
\be
\left ( \frac{t_1}{t_2} \right )^{1-\frac{1}{k N}} \approx \frac{k}{c N}.
\ee
Since the above result is expected to hold for an arbitrary $t_1$ and $t_2$, we conclude that
\be
\chi(t) = k t = \frac{t}{N}
\label{eq:cross-correlation}
\ee
and
\be
\xi(t) = c t = \frac{t}{N^2}.
\label{eq:coherence}
\ee
Moreover Eq. (\ref{eq:small_loops_decay}) becomes 
\be
\frac{d L(t)}{dt} = -\frac{L(t)}{t}.
\label{eq:small_loops_decay2}
\ee
with solution $L(t) \propto 1/t$.

Before we move on let us quickly describe an equivalent model which we call the "dating model". Consider a very long lineup of $L$  men and $L$  women moving against each other looking to marry someone of his (her) own type under assumption that there are only $N$ distinct types of men and women. Once this happens a happy couple is removed from the lineup (they get married, have kids and never go back to dating). It is a simple numerical exercise to check that the evolution of the lineup is describe by equations (\ref{eq:cross-correlation}), (\ref{eq:coherence}) and (\ref{eq:small_loops_decay2}). If initially men and women are distributed randomly, then at the very first step the probability to find the right person is $1/N$, but in the long run the probability goes down because of cross-correlations.  So if someone did not get married by some late time, then he will be surrounded by "friends" that  share the same type and gender, but when he does get married all of his unmarried friends are likely to get married as well. In the language of strings we are likely to see very large bursts of small loops moving with large velocities in the same direction as was previously argued in \cite{PR, Vanchurin}. Of course the dating model does not describe all of the important dynamics on all scales, but nevertheless it provides us with a very simple and intuitive picture of non-linear dynamics. All of the additional effects such as cosmological stretching and back-reaction of loops will be analyzed in the following section. 

\section{Evolution of power}

A key assumption of the previous section was that the number of directions $N$ does not change in time. Since this number is closely related to the average amplitude of wiggles $P(l_{min}) \sim 1/\sqrt{N}$ we must check that $P(l_{min},t)$ does not vary too fast. There are two main mechanism which could lead to the evolution of power on the smallest scales:  emission of loops and cosmological stretching. 

The first mechanism is caused by an enhanced production of small loops in the vicinity of cusps.  Any left-moving wiggle meets $N(t)^2 2 dt/t$ distinct right-moving wiggles per time $dt$, but because of the cross-correlations only $2 N(t) dt/t \sim 2 P(t)^{-2} dt/t$ of them are pointing in a random directions. Moreover the probability to have a significant overlap, in which case a significant part of the wiggle will be removed, is proportional to $P(t)^2$.   As a result, the average amplitude on scales of $l_{min}$ will be decreasing according to
\be
\frac{d P(t)^2}{dt} \propto -\frac{P(t)^4}{t}  ,
\ee
or
\be
\frac{d P(t)}{dt} \propto - \frac{P(t)^3}{t}.
\label{eq:emission}
\ee
In other words the wiggles with larger amplitudes are more likely to form cusps, and therefore loops, which results in an overall decrease of power on the scales comparable to the loop sizes.

The second mechanism that might be responsible for the reduction of power is cosmological stretching \cite{CK2} which is described by the following equation
\be
P(t_2) = P(t_1) \left ( \frac{a(t_2)}{a(t_1)}\right )^{2 \langle v^2 \rangle - 1},
\label{eq:stretching0}
\ee
where $a(t) \propto  t^\alpha$ is the scale factor and $\langle v^2 \rangle$ is the mean square velocity of strings.  If we neglect the cross-correlations between opposite moving waves $ \langle \bold{a}'(\sigma) \bold{b}'(\sigma)  \rangle = 0$, then $\langle v^2 \rangle$  would be $1/2$, but as we now know production of small loops introduces significant cross-correlations. Such cross-correlations can be roughly estimated as
\be
2 \langle v^2 \rangle  -1 = - \langle \bold{a}'(\sigma) \bold{b}'(\sigma)  \rangle \sim  -\frac{1}{N(t)} \sim -P(t)^2
\label{eq:mean_velocity}
\ee
This is because the opposite moving segments can point in all directions but one. By combining  (\ref{eq:stretching0}) and (\ref{eq:mean_velocity}) we arrive to a simple differential equation which describes stretching of wiggles on cosmological backgrounds
\be
\frac{d P(t)}{dt} \propto -\alpha \frac{P(t)^3}{t},
\label{eq:stretching2}
\ee
and has a from identical to (\ref{eq:emission}).  

It follows that a complete evolution equation with both effects (stretching and emission of loops) taken into account is obtained by combining (\ref{eq:stretching2}) and (\ref{eq:emission}): 
\be
\frac{d P(t)}{dt} = -\left (A + \alpha B \right ) \frac{P(t)^3}{t}.
\label{eq:combined_equation}
\ee
where $A$ and $B$ are some constants of order one. The corresponding solution is
\be
P(t)  =  (C + 2 \left (A + \alpha B \right )  \log(t))^{-\frac{1}{2}} , 
\ee
or
\be
N(t)  =  C + 2 \left (A + \alpha B \right )  \log(t).
\label{eq:logarithmic}
\ee
This shows that the number of directions grows only logarithmically which is self-consistent with our original assumptions. 

We are now ready to understand the non-scaling behavior observed in many numerical simulations. At the initial conditions the consecutive segments of size $l_{min}$ have large angles $\sim \pi/4$ which makes them an easy target. The corresponding $N \sim 10$ and the relevant length scales grow as 
\be
\chi \sim 0.1 t
\ee
and 
\be
\xi \sim  0.01 t
\ee
according to (\ref{eq:cross-correlation}) and (\ref{eq:coherence}). We would like to note that at early times the three length scales:  correlation length $\zeta$, inter-string distance $d$ and cross-correlation length $\chi$ are approximately of the same order, but the coherence length is one order of magnitude smaller. Therefore, throughout the evolution one can neglect possible intersections of the nearby strings and production of large loops which would not significantly modify correlations and cross-correlations established on the string. In this limit our non-scaling model can describe fairly well the non-scaling evolution of wiggles and production of small loops. 

In general the decay of string length in a scaling network must be given by 
\be
\frac{d L(t)}{dt} = -\Gamma_{small} \frac{L(t)}{t}-\Gamma_{large} \frac{L(t)}{t}-\Gamma_{friction} \frac{L(t)}{t}.
\label{eq:overall_decay}
\ee
with $\Gamma_{small} + \Gamma_{large}  + \Gamma_{friction} = 2$. The three terms describe the transfer of energy to small loops, large loops and Hubble friction respectively. From (\ref{eq:small_loops_decay2}) the decay parameter of small loops is exactly one, but because of the logarithmic growth (\ref{eq:logarithmic}) it might be a bit smaller $\Gamma_{small} \sim 1$. Since the decay rate of the small scale power due to stretching  \cite{book} is described by the mean velocity of strings (\ref{eq:mean_velocity}), in the presence of very strong cross-correlations predicted by our non-scaling model we get
\be
\Gamma_{friction}  = - \alpha (2 \langle v^2 \rangle  -1) \sim \alpha P(t)^2 \ll 1.
\ee
which is already subdominant at early times and would decay logarithmically as more structures on small scales are eliminated.  Therefore our analysis suggests that in addition to cosmological stretching and non-scaling small loops we must have a population of intermediate or large loops with $\Gamma_{large}  \sim 1$ which will be analyzed extensively in a forthcoming publication \cite{Vanchurin3}.

\section{Conclusions}

The main objective of this paper was to construct a dynamical model of cosmic strings with back-reaction from non-scaling loops taken into account. This was largely motivated by analytical  results obtained in \cite{DPR, PR} by perturbative methods which cannot be used to study the production of loops at larger scales $\gg l_{min}$. The model is far from being complete, as it does not include all of the important effects on the intermediate and large scales, but is mature enough to make a number of interesting predictions. First of all the model predicts an existence of two new length scales: the coherence length and the cross-correlation length. Secondly, the model shows that the small scales power decays only logarithmically and as a result the smallest scales must play a key role in the dynamics of cosmic strings. 

Throughout the paper we have  explicitly  assumed that the gravitational back-reaction scale remains always smaller than the scale of the initial conditions $l_{min}$. However, if the back-reaction scale can grow linearly with time (which may or may not be the case) then this assumption is likely to be broken sooner or later.  From that point on, the relevant production scale of small loops $l_{min}$ might also start to scale, but the correlation and cross-correlation properties analyzed in the paper might not be significantly modified given that $l_{min} \ll t$. We expect that the only relevant consequence of this scenario would be that the primary scale of small loops emitted at cusps would also scale with time. On the other hand the logarithmic decay of power at the scales of $l_{min}$ derived in the previous section cannot continue indefinitely and the network must enter full scaling. We plan to address this matter in the future work. 

\section*{Acknowledgments}

I am very grateful to Alex Vilenkin for our numerous discussions in which my ideas where sharpened to their present form.

\end{document}